# ScPd$_2$Al$_3$ – New Polymorphic Phase in Al-Pd-Sc System

Jiří Pospíšil[1,2*], Yoshinori Haga[1], Kunihisa Nakajima[3], Norito Ishikawa[3], Ivana Císařová[4], Naoyuki Tateiwa[1], Etsuji Yamamoto[1], Tomoo Yamamura[5]

[1] *Advanced Science Research Center, Japan Atomic Energy Agency, Tokai, Ibaraki, 319-1195, Japan*
[2] *Charles University in Prague, Faculty of Mathematics and Physics, Department of Condensed Matter Physics, Ke Karlovu 5, 121 16 Prague 2, Czechia*
[3] *Nuclear Science and Engineering Center, Japan Atomic Energy Agency, Tokai, Ibaraki, 319-1195, Japan*
[4] *Department of Inorganic Chemistry, Faculty of Science, Charles University in Prague, Hlavova 8, 128 43 Prague 2, Czechia*
[5] *Institute for Materials Research, Tohoku University, Sendai, 980-8577, Japan*

**Abstract**
We have discovered a new compound of the composition ScPd$_2$Al$_3$ crystallizing in unknown structure type. Moreover, ScPd$_2$Al$_3$ reveals polymorphism. We have found an orthorhombic crystal structure at room temperature and a high temperature cubic phase. The polymorphic phases are separated by a reversible first order transition at 1053°C with a hysteresis of 19°C. ScPd$_2$Al$_3$ exists as a very stable intermetallic phase just in the vicinity of the icosahedral quasicrystal Tsai-type *i*-phase Al$_{54}$Pd$_{30}$Sc$_{16}$.

Keywords: ScPd$_2$Al$_3$, Al-Pd-Sc, polymorphism, YPd$_2$Al$_3$

**Introduction**

Intermetallics of the composition *RE*Pd$_2$Al$_3$ (*RE* – rare earth elements) crystallizing in the hexagonal PrNi$_2$Al$_3$-type structure represent extraordinary diversity of the physical properties. CePd$_2$Al$_3$ exhibits heavy-fermion behavior at low temperatures with a Kondo effect. Nevertheless the properties are strongly sample dependent and also antiferromagnetic order was detected[1-3]. PrPd$_2$Al$_3$ remains paramagnetic down to 1.5 K due to the crystal field effect[4, 5]. NdPd$_2$Al$_3$ becomes antiferromagnetic at $T_N$ = 6.5 K but $T_N$ varies in the off-stoichiometric samples[6]. The most intriguing cases are the frustrated triangular lattice antiferromagnets GdPd$_2$Al$_3$ and SmPd$_2$Al$_3$[7-9], and superconductors (SC) LaPd$_2$Al$_3$ ($T_{SC}$ = 0.85 K)[10] and YPd$_2$Al$_3$ ($T_{SC}$ = 2.2 K)[11]. It is worth noting that this group also covers the unique actinide SCs UPd$_2$Al$_3$[12], UNi$_2$Al$_3$[13] and ThPd$_2$Al$_3$[14]. Recent discovery of the SC in YPd$_2$Al$_3$ with the highest SC transition motivated us to prepare so far unknown analogue with the lightest rare earth element Scandium.

**Experimental part**

The polycrystalline sample of ScPd$_2$Al$_3$ was prepared by a conventional arc melting procedure of the stoichiometric amounts of elements (Sc-3N, Pd-3N5, Al-5N) under Ar protective atmosphere. The sample of mass 2.46 g was re-melted three times to ensure good homogeneity. No sign of evaporation was observed during the melting with mass loss only 4.5 ‰. The sample was annealed at 900°C for 7 days wrapped in Ta foil under vacuum. The crystal structure was studied by single crystal X-ray diffraction using R-AXIS RAPID (Rigaku) (Mo-K$\alpha$-radiation) and X-ray powder diffraction (XRPD) by Rigaku Ultima IV diffractometer (Cu-K$\alpha$-radiation). Electron-probe analyzer EPMA JXA-8900 (JEOL) was used for the chemical analysis. Single crystal diffraction patterns were analyzed by SHELX software[15]. Rietveld method implemented to FullProf software was used to refine the XRPD patterns[16, 17]. Electrical resistivity and heat capacity measurements were carried out by a commercial PPMS 9T Quantum Design device. Differential thermal analysis (DTA) was performed up to 1200°C in the rate 20°C/min under the continuous 6N Ar flow of 140 ml/min using TG/DTA7300 model thermal analyzer from Seiko Instruments Inc. (SII).

**Results and discussion**

$ScPd_2Al_3$ is a silver, highly reflective, and very fragile metallic material. Microprobe analysis of the as-cast and annealed sample confirmed a homogeneous distribution of the elements with no sign of spurious phases. However, we failed to fit the XRPD pattern by the expected hexagonal lattice of the $PrNi_2Al_3$ structure-type. In addition, a series of new reflections appeared in the XRPD pattern of the annealed sample besides the original reflections with reduced intensity see-Fig. 1. Since the microprobe analysis did not reveal any spurious phases, we excluded thermal decomposition. Solution is only a polymorphic transition, which was initialized by annealing procedure and portion of a high temperature (HT) polymorph of different structure was retained in the annealed sample.

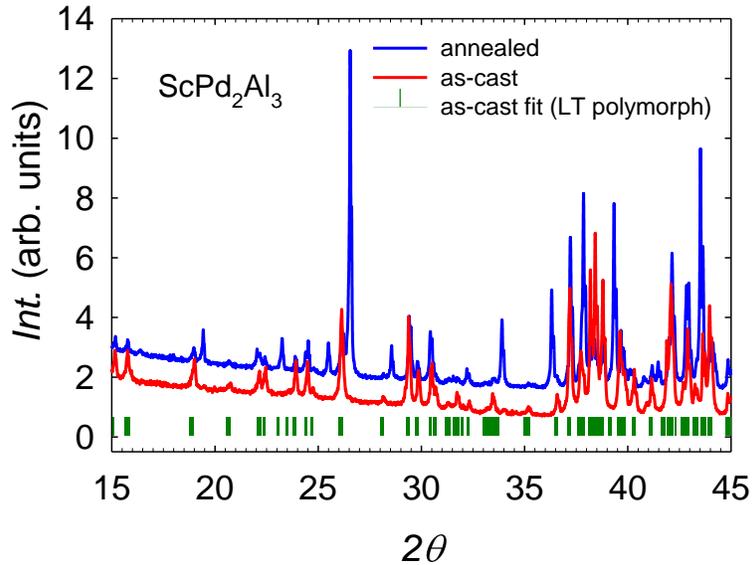

**Fig. 1.** XRPD patterns of the as-cast and annealed sample of $ScPd_2Al_3$. The green ticks represent generated Bragg reflections of the LT polymorph using the structural model received by the single crystal diffraction analysis of the as-cast sample. Series of new reflections appeared in the annealed samples undescribed by LT polymorph crystallizing in the orthorhombic structure.

We used DTA method to study this polymorphic transition. The first thermal cycle using the as-cast sample revealed only one peak in the heating branch at 1072°C. It is clear evidence that the as-cast sample consisted only of a low temperature (LT) polymorph. Otherwise, two characteristic peaks had to appear in the first heating branch in the case of quenched HT polymorph[18-20]. We have estimated transformation temperatures as the edges of the peaks[21]. The transformation between high temperature HT and LT polymorph is characterized by a hysteresis of 19°C because the corresponding anomaly appeared at 1053°C in the cooling branch (Fig. 2). We have carried out second thermal cycle using the identical sample after the first cycle. No apparent difference between both cycles with the conserved integrated intensity of the peaks in the cooling and heating branch supports a fully reversible first order transition. It is also a signature that both polymorphs are thermodynamically very stable without evidence of a decomposition process. Applied cooling rate during the DTA measurement of 20°C/min was not fast enough to quench the HT polymorph to room temperature.

A growth of the HT polymorph was evidently initialized during the annealing process of the sample just below the temperature of polymorphic transformation. In such a special case, termination of the annealing process by rate only ~-1 K/min retained of a portion of the HT polymorph downwards to room temperature. Two-phase sample was harder than the as-cast one and size of the grains was also significantly reduced.

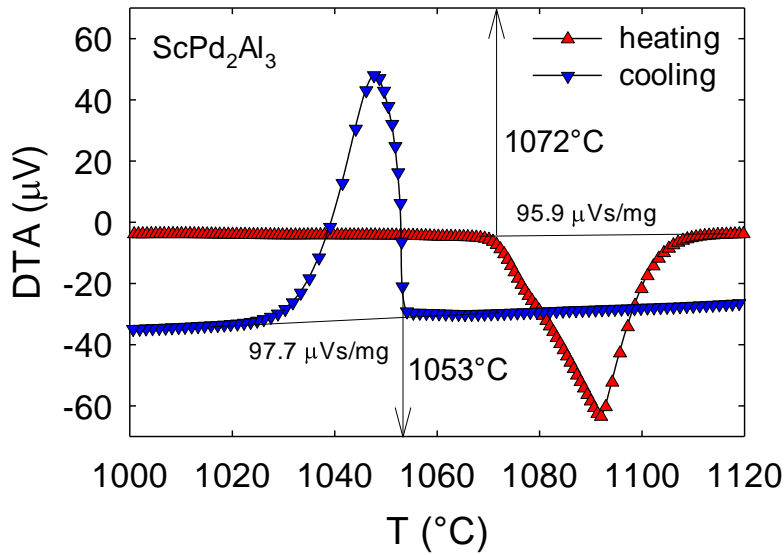

**Fig. 2.** DTA curves of the ScPd$_2$Al$_3$ (second run).

We successfully separated several single crystal grains from the as-cast sample suitable for the single crystal X-ray diffraction. We have always received set of patterns, which can be solved using orthorhombic crystal structure of the large lattice parameters $a = 8.6400(3)$ Å, $b = 22.4192(8)$ Å and $c = 8.5977(3)$ Å of volume $V = 1665.39$ Å$^3$. Extinction rules are consistent with a *C*-centered orthorhombic cell. *Cmmm* is the likely space group based on the intensity distribution. Present unit cell is expected to contain 16 formula units. Generated reflections are in agreement with the experimental XRPD pattern of the as cast sample (Fig. 1). Notwithstanding, we have received high quality patterns with very sharp reflections, final solution can be carried out only by combination with advance methods as an electron diffraction or high resolution scanning transmission microscopy because of crystal structure complexity. The anomalous scattering using specific X-ray wavelength of Sc, Pd and Al absorption edges can solve the uncertainty of the atoms mixture or vacancies in the specific sites [22].

To analyze the HT polymorph, we have utilized fact that the short annealing (1 hour) of the sample in the fine powder form at 1150°C under vacuum followed by a cooling ~-5 K/min led to stabilization of the HT polymorph down to room temperature. We were able to separate several single crystal grains to perform single crystal diffraction. We could successfully index data by a cubic unit cell with the lattice parameter $a = 8.9462(12)$ Å with a corresponding unit cell volume $V = 716.00$ Å$^3$. Identical result was received from all studied grains. However, lower quality of the diffraction patterns of the HT polymorph because of size of the grains allowed only the solution of the lattice parameters.

Polymorphic transition seems to be complex in $ScPd_2Al_3$. Rapid cooling after the melting of the bulk sample in the arc furnace was not sufficient to retain the HT polymorph as a metastable phase at room temperature in agreement with observation by DTA measurement using cooling and heating rate 20 K/min. On the other hand, HT phase was evidently stabilized during the annealing of the powder and partially in the bulk sample when the annealing temperature only approaches the temperature of the polymorphic transition. Cooling rate is typically the key parameter to retain the HT polymorph as a metastable phase at room temperature. However, another parameters like the upper temperature of the annealing[23] and particles dimension can strongly affect the dynamic and kinetics of the polymorphic transition[24-26]. $ScPd_2Al_3$ with complex structure seems to be the case.

Electrical resistivity and heat capacity of the as-cast and annealed samples, where both polymorphs were presented, did not reveal SC transition between the room temperature and 2 K.

We also discuss the origin of the deviation of the $ScPd_2Al_3$ from the hexagonal lattice. Since the size of ionic radius 0.0703 nm of $Sc^{3+}$ is quite smaller than those of other trivalent rare-earth ions, e.g. 0.095 nm of La, 0.0905 nm of Y, the crystal structure of $ScPd_2Al_3$ may be different from other $RE$Pd$_2$Al$_3$ [27, 28]. The ionic radius of $Sc^{3+}$ is smaller than those of heavier rare-earth ions. The change of crystal structure of rare-earth intermetallic compounds is sometimes seen at the boundary of lighter rare-earth and heavier rare-earth compounds [29].

Phase diagram of the Al-Pd-Sc system is frequently studied because $Al_{54}Pd_{31}Sc_{15}$ is a well-known 1/1-crystal approximant to the icosahedral quasicrystal Tsai-type $i$-phase of the composition $Al_{54}Pd_{30}Sc_{16}$[30, 31]. Detailed study has shown that the approximant phase and $i$-phase exist in the extremely narrow region of the Al-Pd-Sc phase diagram. The $i$-phase transforms to other phase(s) via an exothermic reaction at 1098 K (825°C), pointing that this phase is metastable[30]. This temperature is significantly lower than that of our case. The re-scaled composition of the $i$-phase is $ScPd_{1.875}Al_{3.375}$, which is Al richer and Pd poorer in comparison with $ScPd_2Al_3$.

**Conclusions**

Although majority of the $RE$Pd$_2$Al$_3$ compounds keep hexagonal PrNi$_2$Al$_3$-type structure, we have recognized two exceptions. EuPd$_2$Al$_3$[32], and the newly discovered $ScPd_2Al_3$, which exists just in the proximity of the icosahedral quasicrystal Tsai-type i-phase $Al_{54}Pd_{30}Sc_{16}$ and its approximant. Deviation of the $ScPd_2Al_3$ from the hexagonal PrNi$_2$Al$_3$-type structure most likely originates in $Sc^{3+}$ ionic diameter. $ScPd_2Al_3$ is a stable intermetallic phase with a polymorphic phase transition at 1053°C with hysteresis 19°C. Orthorhombic and cubic crystal structure was solved for the LT and HT polymorph, respectively. Solution of the complex crystal structures as well as the annealing process to receive the metastable HT polymorph at room temperature will be subject of the further research by a splat cooling method.

**Acknowledgements**
This work was supported by the JSPS KAKENHI No. JP15H05852, 16K05463, and 15H05884 (J-Physics).

*jiri.pospisil@centrum.cz